\documentclass[aps,prl,reprint,showkeys]{revtex4-2}
\pdfoutput=1
\usepackage{amsmath}
\usepackage{amssymb}
\usepackage{amsfonts}
\usepackage{graphicx}
\usepackage{bm}
\usepackage{color}

\renewcommand\vec{\mathbf}
\def\uGrad{\vec{\nabla}}

\newcommand{\ud}{\,{\mathrm{d}}}

\def\uiiint{\int\!\!\!\int\!\!\!\int}

\def\uvB{\vec{B}} 
\def\uC{C} 
\def\uD{D} 
\def\ue{e} 
\def\uE{E} 
\def\ueE{\epsilon} 
\def\ueEDM{\epsilon_\mathrm{DM}} 

\def\ueEMS{\epsilon_\mathrm{MS}} 
\def\ugammaB{\gamma_\mathrm{B}}
\def\uHopf{H} 
\def\uh{h} 
\def\uH{H_\mathrm{Z}} 
\def\uvH{\vec{H}}     
\def\uvHD{\vec{H}_\mathrm{D}} 
\def\uK{K}         
\def\ulambda{\lambda}
\def\uLd{L_\mathrm{D}}
\def\uLe{L_\mathrm{E}}

\newcommand{\uvM}{\vec{M}}
\def\uMi{M_i}      
\newcommand{\uvm}{\vec{m}}
\def\umi{m_i}      
\newcommand{\umx}{m_{\mathrm{X}}}
\newcommand{\umy}{m_{\mathrm{Y}}}
\newcommand{\umz}{m_{\mathrm{Z}}}
\def\uO{O}         
\def\uP{P_i}
\def\uPex{P_\mathrm{EX}}
\def\uPa{P_\mathrm{A}}
\def\uPz{P_\mathrm{Z}}
\def\uPdm{P_\mathrm{DM}}
\def\uPms{P_\mathrm{MS}}
\def\up{p_i}
\def\upex{p_\mathrm{EX}}
\def\upa{p_\mathrm{A}}
\def\upz{p_\mathrm{Z}}
\def\updm{p_\mathrm{DM}}

\newcommand{\uMs}{M_{\mathrm{S}}}
\newcommand{\uMssq}{M_{\mathrm{S}}^2}
\def\umuZ{\mu_0}
\def\uq{q} 
\def\uvr{\vec{r}}          
\def\uvrp{\widetilde{\vec{r}}}  
\def\uvs{\vec{s}}  
\def\uX{X}         
\def\uY{Y}         
\def\uZ{Z}         
\def\uXp{\widetilde{X}} 
\def\uYp{\widetilde{Y}} 
\def\uZp{\widetilde{Z}} 
\graphicspath{ {./figures/} }

\begin{document}
\title{Two types of metastable magnetic hopfions}

\author{Konstantin L. Metlov}\email[Electronic address: ]{metlov@fti.dn.ua}
\affiliation{Donetsk Institute for Physics and Technology, R.~Luxembourg str.~72, 83114~Donetsk, Ukraine}
\affiliation{Institute for Numerical Mathematics RAS, 8~Gubkina str., 119991~Moscow GSP-1, Russia}

\date{\today}

\begin{abstract}
Localized magnetic topological solitons with Hopf index of 1 in an unbounded bulk magnet are studied theoretically, starting with the classical micromagnetic Hamiltonian. It is shown analytically that (like Bloch and N{\'e}el walls in classical magnetism) there are two possible stable configurations of these solitons, which are equivalent topologically. Their stability regions are plotted on a phase diagram in the magnetic field -- uniaxial anisotropy space for different normalized magnetostatic interaction strengths. For some parameters both types of hopfions can be stable.
\end{abstract}


\maketitle
Topologically non-trivial objects are at the core of the magnetism: domain walls~\cite{Hubert_book_walls}, Bloch lines~\cite{Slonczewski1974}, magnetic bubbles~\cite{bobeck1975bubbles}, skyrmions~\cite{BY1989,*ISZ90,BH1994}, magnetic vortices and antivortices in planar nanoelements~\cite{M10} are some of many examples of magnetic solitons~\cite{KIK90} in 1 and 2 spatial dimensions. Recent advances in 3-dimensional (3D) imaging techniques~\cite{DGSGHRH2017} open new possibilities to look for magnetic solitons in 3D -- like Bloch points~\cite{DGSGHRH2017}, magnetic vortex rings~\cite{CMSGHBRHCG2020} or the pinned hopfion states~\cite{KRRCVDCHSPFHSF2021} in multilayer thin film elements. There are theoretical examples of topologically non-trivial configurations in bulk magnets with complex/unusual exchange interactions~\cite{Nicole1978,RKBDMB2019}, but the case of bulk 3D solitons in systems, described by classical micromagnetics Hamiltonian with long-range dipolar interactions, remains largely unexplored.

Consider a magnetic material, modeled as a 3D continuum with a spatial distribution of local magnetization vectors $\uvM(\uvr)$ having a fixed (constant in space, but temperature-dependent) magnitude $|\uvM|=\uMs>0$. Non-zero value of $\uMs$ (below of a certain critical temperature) is due to quantum-mechanical exchange interaction between the spin carriers in the material dominating over the temperature fluctuations. Ferromagnetic exchange interaction tries to keep the neighbouring magnetic moments aligned and introduces a steep energy cost for fast variations of $\uvM(\uvr)$. Magnetization discontinuities have infinite exchange energy in the continuum approximation. In reality, owing to the ultimate discreteness of the material and the presence of defects, the localized singularities of the magnetization vector field --- the Bloch points~\cite{Feldtkeller1965,*doring68} --- are also possible. But in the present work we will focus only on the continuous vector fields with defined topology.

Boundary conditions are of the key importance for observing the topological effects. Just like it is with knots on a rope, having both ends of the rope connected (e.g. by holding them in hands) allows to introduce their topologically non-trivial classes. But all of them can be trivially untied if one end of the rope comes loose (or the rope is cut, as in the Gordian knot legend). Here we will require that the magnetization somewhere infinitely far away (no matter the direction) has a particular orientation. Such a condition is natural for an infinite magnet in a uniform external magnetic field. Mathematically it means that there is a single infinitely distant point or that the volume of the considered 3D magnet can be mapped to a sphere $S^3$ in four dimensions. Given that the ends of the magnetization vectors of fixed magnitude lie on a three-dimensional sphere $S^2$, the singularity-free magnetization distribution in an infinite 3D magnet with a single ``infinity'' point can be understood as a continuous mapping $S^3\rightarrow S^2$.

For a long time in mathematics it was accepted that all such mappings are homotopically equivalent until Hopf~\cite{Hopf1931} provided a counterexample, which was later generalized by Whitehead~\cite{whitehead1947}. It turned out that all $S^3\rightarrow S^2$ mappings can be subdivided into homotopy classes numbered by an integer $\uHopf$, called the Hopf index (or the Hopf invariant). Topologically trivial mappings correspond to $\uHopf=0$, the original example~\cite{Hopf1931} by Hopf has $\uHopf=1$ and Whitehead gave the expression~\cite{whitehead1947} for mappings with an arbitrary $\uHopf$.

In physics, however, the topology is not everything. Take, for example, 1D magnetic domain walls (corresponding, with a similar boundary conditions, to the maps $S^1\rightarrow S^2$), which can be of Bloch or N{\'e}el type. Both these types of walls are topologically equivalent. Yet, they have very different spatial profiles, different dynamics and different stability regions on the magnetic phase diagram. They are different objects, which even may coexist in some circumstances~\cite{LCCSC98}.

Not only there can be several different (but topologically equivalent) stable magnetization configurations, some topologically non-trivial configurations can also be unstable. This might seem like a paradox: if one can not untie a knot without introducing discontinuities and overcoming the corresponding (infinite) exchange energy barriers, how a topologically non-trivial configuration can be destroyed ? The answer is simple --- by rescaling. The knot is only stable if it has a well-defined scale (e.g. its enlargemnent or shrinkage leads to the energy increase), otherwise it will either collapse into a point or expand into infinity and this way effectively disappear. Thus, the energetic aspects of knot stability are just as important as topological ones.

Let us start with the following well known expression for the energy density of a magnet:
\begin{eqnarray}
        \ue & = & \frac{\uC}{2\uMssq}\sum_{i=\uX,\uY,\uZ} \left|\uGrad\uMi\right|^2 +\frac{\uD}{\uMssq}\uvM\cdot[\uGrad\times\uvM] -\nonumber\\
        & &   -\umuZ \left(\uvM \cdot \uvH\right) - \frac{\umuZ}{2}\left(\uvM \cdot \uvHD\right)- \frac{\uK}{\uMssq}\left(\uvM \cdot \uvs\right)^2,\label{eq:energy}
\end{eqnarray}
consisting of the exchange energy with the exchange stiffness parameter $C=2A$, the chiral Dzyaloshinskii-Moriya (DM) interaction term~\cite{BS1970,*BJ1980} with the corresponding energy constant $D$, the energy of the magnetic moment in the external magnetic field $\uvH$, the magnetostatic energy of the magnetic moment in the demagnetizing field $\uvHD$, created by all the magnetic moments in the material as per Maxwell's equations, and the uniaxial anisotropy energy with the constant $\uK$ and the axis director $\uvs$. Following Aharoni~\cite{AharoniBook}, we will use the relation  $\uvB = \umuZ(\uvH + \ugammaB \uvM)$ for the magnetic induction to cover all common systems of magnetic units [$\umuZ$ is the permeability of vacuum and  $\ugammaB=1$ in SI; $\ugammaB=4\pi$, $\umuZ=1$ in CGS].

Let us choose the Cartesian coordinate system in such a way that the anisotropy axis is directed along the $\uO\uZ$ ($\uvs=\{0,0,1\}$) and restrict ourselves to the case when the external field $\uvH=\{0,0,\uH\}$ is also parallel to $\uO\uZ$. The problem of finding the stable magnetization configurations, corresponding to the minimum of the total energy $\uE=\uiiint\ue\ud^3\uvr$ with the density~\eqref{eq:energy}, has two characteristic length scales: the exchange length $\uLe=\sqrt{\uC/(\umuZ\ugammaB\uMssq)}$ and the Dzyaloshinskii-Moriya length $\uLd=\uD/(\umuZ\ugammaB\uMssq)$. Just like in the theory of periodic lattice of skyrmions~\cite{BY1989}, it is convenient to renormalize all quantities to $\uD$ by working with the normalized dimensionless total energy $\ueE=\uE\uC/\uD^2$ per volume of the hopfion lattice unit cell, introducing the dimensionless external field $\uh=\umuZ\uMs\uC\uH/\uD^2$, the dimensionless analog of the anisotropy quality factor $\uq=2\uC\uK/\uD^2$ and the dimensionless parameter $\ulambda=\uLd/\uLe=\uD/\sqrt{\umuZ\ugammaB\uC\uMssq}$, characterizing the relative strength of the DM interaction with respect to the dipolar magnetostatic one.

Now let us look for stable three-dimensional localized magnetization configurations with non-trivial Hopf index, minimizing the total energy of the magnet $\uE$ with specific values of $\uh$, $\uq$ and $\ulambda$. Unfortunately, this problem leads to a system of non-linear partial differential integral equations, which are hard (if not impossible) to solve analytically. It is also known, that such a system usually has many different metastable solutions. Finding a topologically non-trivial one among them numerically requires either the prior (and quite precise) knowledge of the configuration one is looking for or a great deal of luck.

Here we will follow the usual Ritz method, employed in analytical micromagnetics. It is based on selecting a suitable trial function for the magnetization configuration and minimizing the total energy $\uE$ by varying the unknowns in that function. A good starting point for topologically non-trivial configurations is the Whitehead's ansatz~\cite{whitehead1947}, which can be generalized as follows:
\begin{eqnarray}
 \label{eq:R3S3ster}
 &u=\frac{2(\uXp+\imath \uYp)R}{\uXp^2+\uYp^2+\uZp^2+R^2},
 \quad
 v=\frac{R^2-\uXp^2-\uYp^2-\uZp^2+\imath 2 \uZp R}{\uXp^2+\uYp^2+\uZp^2+R^2},
\\
 \label{eq:Whitehead}
 &w=e^{\imath\chi}u^n/v^m, \\
 \label{eq:cplxToM}
 &\{\umx+\imath\umy,\umz\}=\{2w, 1 - |w|^2\}/(1+|w|^2).
\end{eqnarray}
The first equation above is the stereographic projection of the Cartesian coordinates $\uvrp=\{\uXp,\uYp,\uZp\}$ onto a sphere $S^3$, parametrized by two complex coordinates $u$ and $v$ such that $|u|^2+|v^2|=1$; the second is the modified Whitehead's ansatz, mapping the sphere $S^3$ onto a sphere $S^2$, parametrized by the complex coordinate $w$; finally, the last equation is a stereographic projection, mapping the sphere $S^2$ onto the components of the reduced magnetization vectors $\uvm=\uvM/\uMs=\{\umx,\umy,\umz\}$, such that $|\uvm|=1$. The parameter $R$ is the (yet unknown) scale factor, the parameter $\chi$ is a free angle, allowing to rotate the target sphere $S^2$. When $n$ and $m$ are mutually prime integers, the above ansatz generates a vector field $\uvm(\uvrp)$ such that the pre-image of every particular magnetization direction $\uvm_1$ is a closed curve in $\uvrp$, which is interlinked with a curve, corresponding to any other magnetization direction $\uvm_2\neq\uvm_1$, exactly $m n$ times~\cite{whitehead1947}. In other words, the vector field $\uvm(\uvrp)$ is topologically non-trivial with the Hopf index $\uHopf=m n$. Only the simplest case $\uHopf=m=n=1$ is considered here, which can also be expected to have a lower total energy $\uE$, compared to the magnetization configurations with higher Hopf index. In the center of such a hopfion $\uvrp=0$ and at the infinity $|\uvrp|\rightarrow\infty$ the magnetization has a fixed direction $\uvm=\{0,0,1\}$.

While the above ansatz sets the topology of the magnetization distribution, its scale $R$ is not yet defined. There is also a freedom of selecting a smooth one-to-one map between the physical space $\uvr$ and the space $\uvrp$, which will preserve the topological properties of the vector field. Because it can be {\em a priori} expected that 3D magnetic solitons are localized in space (otherwise their magnetic energy will be infinite), let us enforce their strong localization within a sphere of the radius $R$, by introducing the map
\begin{eqnarray}
 \label{eq:StoR}
 &\{\uXp,\uYp,\uZp\}=\frac{\{\uX,\uY,\uZ\}}{1-f(\sqrt{\uX^2+\uY^2+\uZ^2}/R)},
\end{eqnarray}
where $\uX^2+\uY^2+\uZ^2<R^2$ and the function $f(x)$, $0<x<1$ is monotonous $f'(x)\ge0$ and satisfies the boundary conditions $f(0)=0$, $f'(0)=0$, $f(1)=1$. Such solitons have $\uvm=\{0,0,1\}$ in their center $\uvr=\{0,0,\uZ\}$ and on the surface $|\uvr|=R$, which is mapped into the infinitely distant point in the $\uvrp$ space. Monotonicity is required for the map to be one-to-one and the condition $f'(0)=0$ ensures that the map is smooth at the origin.

Under the above assumptions, the problem of finding stable topologically non-trivial $\uHopf=1$ hopfions reduces to finding the unknown function $f(x)$ and the unknown parameters $R$ and $\chi$, minimizing the total magnetic energy $\uE$ of the magnetization distribution $\uvm(\uvr)$. This problem is much more manageable.

\begin{figure*}
\begin{center}
 \includegraphics[width=\textwidth]{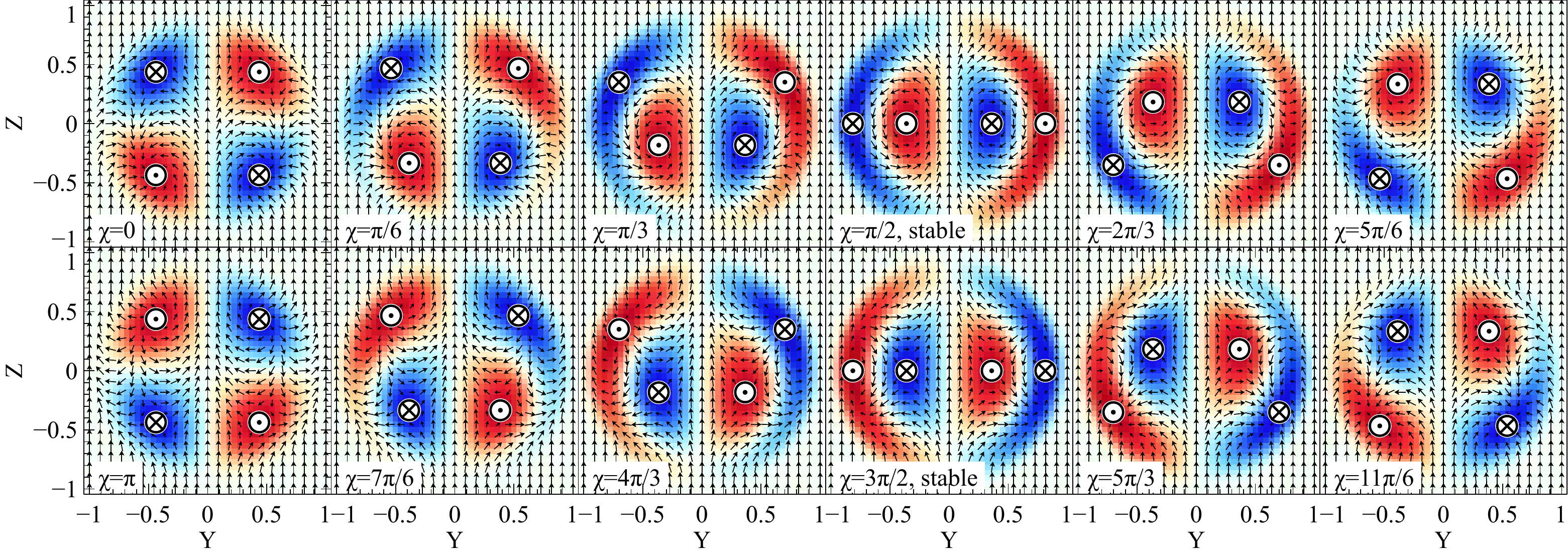}
\end{center}
\caption{\label{fig:chiHopfion}Cross-section of the $\uHopf=1$ hopfion~\eqref{eq:R3S3ster}--\eqref{eq:StoR} with $R=1$ and $f(x)=x^2$ by a plane, passing through its axis of symmetry, for different values of $\chi$ in~\eqref{eq:Whitehead}. The arrows show the projection of the normalized magnetization vectors onto the cutting plane, while the shading (marked by out-of-plane arrows) shows the out-of-plane magnetization component. Only the $\chi=\pi/2$ and $\chi=3\pi/2$ hopfions are stable.}
\end{figure*}
The dependence of the hopfion structure on $\chi$ corresponds to the rotation of the vortex around antivortex in each of the hopfion's axial cross sections, as shown in Fig.~\ref{fig:chiHopfion}. The $\chi$ dependence of the total energy is particularly simple and can be minimized analytically. Really, the total exchange, Zeeman and the anisotropy energies of the $\uHopf=1$ hopfion~\eqref{eq:R3S3ster}--\eqref{eq:StoR} do not depend on $\chi$ at all. This is because the change of $\chi$ implies a uniform rotation of all the spins around $\uO\uZ$ axis, coinciding with the direction of the external field and the anisotropy. The dependence of the total DM energy on $\chi$ has the form $\ueEDM={\cal A}+{\cal B}\sin\chi$, as one can see from the direct computation (also given in the supplemental Mathematica notebook~\cite{suppmathfile}). As $\chi$ takes values from $0$ to $2\pi$, the DM energy has exactly one minimum and one maximum at $\chi=\pi/2$ and $\chi=3\pi/2$. However, the DM interaction constant $\uD$, which is a prefactor to $\ueEDM$, can be both positive and negative (there is also an additional $\pm$ sign, depending on the relative chirality of the hopfion and of the material). This means that, depending on the sign of $\uD$ (or the relative chirality), either of $\chi=\pi/2$ and $\chi=3\pi/2$ can be an energy minimum.

The magnetostatic energy can be expressed as a volume integral of the density of magnetic charges $\rho=(\nabla\cdot\uvm)$:
\begin{equation}
\label{eq:uPms}
\ueEMS\propto\uPms=\frac{1}{8\pi R^3}\uiiint_{V}\ud^3\uvr\uiiint_{V}\ud^3\uvr^\prime\frac{\rho(\uvr)\rho(\uvr^\prime)}{\left|\uvr -\uvr^\prime\right|},
\end{equation}
where the integration goes over the ball of the radius $R$. This integral can be factored, using the identity
\begin{equation}
\label{eq:sphericalHarmonics}
 \frac{1}{\left|\uvr -\uvr^\prime\right|}=4\pi\sum\limits_{l=0}^{\infty}\frac{1}{2l+1}\frac{r_\mathrm{min}^l}{r_\mathrm{max}^{l+1}}\sum\limits_{m=-\infty}^{\infty}Y_l^m(\theta,\varphi)\overline{Y_l^m(\theta^\prime,\varphi^\prime)},
\end{equation}
where the spherical coordinates $\uvr=\{r,\theta,\varphi\}$, $\uvr^\prime=\{r^\prime,\theta^\prime,\varphi^\prime\}$ are used, $r_\mathrm{min}=\min(r,r^\prime)$, $r_\mathrm{max}=\max(r,r^\prime)$, $Y_l^m(\theta,\varphi)$ are the spherical harmonics and the overline denotes the complex conjugate. As a function of $\chi$, the density of charges has the form $\rho=\rho_0+\rho_\mathrm{C}\cos\chi + \rho_\mathrm{S}\sin\chi$. The magnetostatic energy~\eqref{eq:uPms} is therefore quadratic in $\cos\chi$ and $\sin\chi$. However, when integrated over the ball or the radius $R$, it can be shown using the expansion~\eqref{eq:sphericalHarmonics} that the coefficient before the linear in $\cos\chi$ term (including $\cos\chi\sin\chi$ as well) vanishes. The $m$ summation includes only the $m=0$ term due to the axial symmetry and all the products of integrals over $\theta$ and $\theta^\prime$ vanish for all $l$ with an arbitrary $f(x)$ (also demonstrated in the Supplemental Mathematica file~\cite{suppmathfile} via direct computation). Therefore, the magnetostatic energy has the expansion $\ueEMS={\cal C} + {\cal D}\sin\chi+{\cal E}\cos^2\chi + {\cal F}\sin^2\chi$, which always has stationary points at $\chi=\pi/2$ and $\chi=3\pi/2$. This is also evident directly from Fig.~\ref{fig:chiHopfion}. One can see that either increase or decrease of $\chi$ from these two stationary values leads to the magnetization configurations, different only by mirroring the spatial coordinates (but not the magnetization vectors) about the $\uZ=0$ plane. Such mirroring of the magnetic charges distribution has no effect on their magnetostatic energy, which is therefore equal for both increased and decreased $\chi$.

The magnetization in these two $\uHopf=1$ hopfion configurations (marked ``stable'' in Fig.~\ref{fig:chiHopfion}) is a double ring with circular vortex and anti-vortex filaments on top of each other. The type 1 hopfion has the inner vortex and an outer anti-vortex filament, whereas in the type 2 hopfion the order of filaments is reversed.

Provided the axial symmetry is preserved (so that the vortices and antivortices in Fig.~\ref{fig:chiHopfion} can't cross the $\uY=0$ line), the topological stability of these configurations is evident, because the 2D topological charge in each half of the cross section is exactly $1$ and can't be changed by any smooth magnetization variation. It remains to show that the hopfions of either of the two types are stable with respect to rescaling. This problem can be formulated in terms of minimizing the dimensionless total energy per unit of volume
\begin{eqnarray}
 \epsilon& = &\frac{1}{4\sqrt{2}}\Big(\nu^2\uPex+\nu\uPdm+\frac{q}{2}(\uPa-4\sqrt{2})+\nonumber\\
 \label{eq:energyDimensionless}
 & & +h(\uPz-4\sqrt{2})+\frac{1}{\lambda^2}\uPms\Big),
\end{eqnarray}
where the numbers $4\sqrt{2}$ come from assuming that the hopfions form a closely packed 3D lattice (hexagonal or cubic) and the energy is computed per volume of its unit cell, the dimensionless parameter $\nu=\uC/(\uD R)$ measures the inverse radius of the hopfion and
\begin{equation}
\uP=\int\limits_0^1\ud x \left[\int\limits_0^{2\pi}\ud \varphi\int\limits_0^{\pi}\ud \theta\, \up\, x^2\sin\theta \right]
\end{equation}
with $\uvr=R\,x\,\{\cos\varphi\sin\theta, \sin\varphi\sin\theta, \cos\theta\}$, $\upex=(R^2/2) \sum_{i=\uX,\uY,\uZ}|\uGrad\umi|^2$, $\updm=R\, \uvm\cdot[\uGrad\times\uvm]$, $\upa=1-\umz^2$, $\upz=1-\umz$ and $\uPms$ is the magnetostatic self-energy defined by~\eqref{eq:uPms}. The dipolar interaction between different hopfions in the lattice is neglected here for simplicity (it would lift the degeneracy between the hexagonal and cubic close packed lattices and may also induce some modifications of the hopfion profile). The integration over the angular variables $\varphi$, $\theta$ in $\uP$ can be done analytically for both hopfion types, and the expressions with the remaining integral over $x$ are the functionals of the unknown function $f(x)$, their explicit form is given in~\cite{suppmathfile}. Minimization of the energy functional~\eqref{eq:energyDimensionless} can be done via solving the corresponding Euler-Lagrange equation for the unknown parameter $\nu$ and the unknown function $f(x)$, which is monotonous $f^\prime(x)>0$ and subject to the boundary conditions: $f(0)=0$, $f'(0)=0$, $f(1)=1$. In the simplest case of negligible magnetostatic contribution ($\lambda\rightarrow\infty$) it is an ordinary differential equation, which can be solved for $\nu$ numerically using the method of shooting. The magnetostatic contribution adds an integral term. It was accounted here iteratively, computing the interaction of the magnetic charges in the present iteration of $f(x)$ with the field, created by the magnetic charges of the previous iteration, until a stationary point is reached (which corresponds to the interaction of the hopfion's magnetic charges with themselves). To seed the iterations a $\lambda\rightarrow\infty$ solution can be used. This procedure is implemented for both hopfion types in the attached Wolfram Mathematica 12 notebook~\cite{suppmathfile}.

\begin{figure}[tb]
\begin{center}
 \includegraphics[width=\columnwidth]{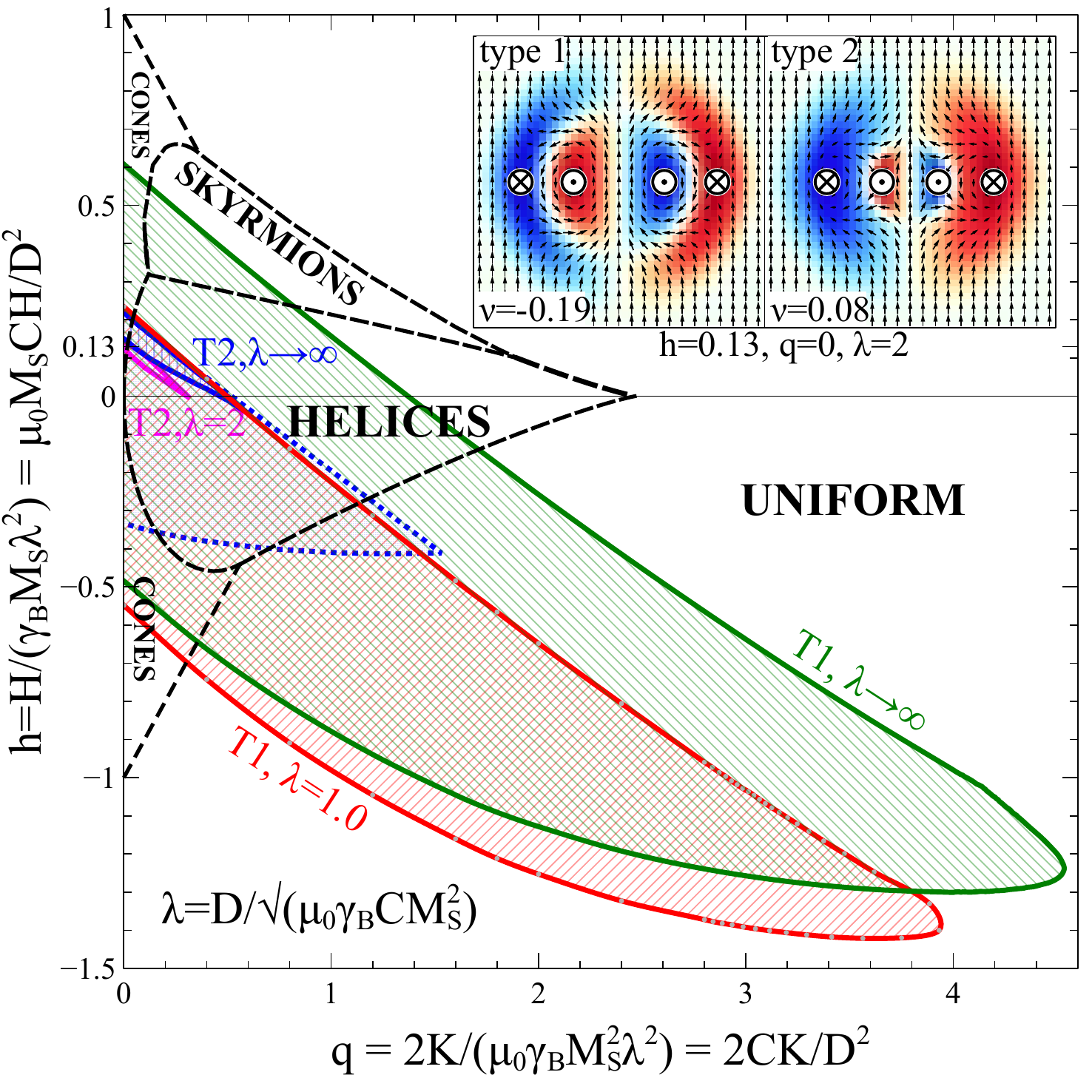}
\end{center}
\caption{\label{fig:phaseDiagram}Map of the ground~\cite{BH1994} and metastable hopfion states of the helimagnet~\eqref{eq:energy} in the reduced field--anisotropy ($\uh$--$\uq$) space for several different values of the relative magnetostatic interaction strength $\lambda$. Dashed lines separate the ground states, marked in capital letters. Solid lines outline the metastability region of type 1 (T1) and type 2 (T2) hopfions for several different values of $\lambda$. The inset shows physical space layout of the T1 and T2 hopfions at a particular point of the map, where both of them are metastable. The dotted contour outlines the stability region of the T2 hopfions at $\lambda\rightarrow\infty$ without requiring the monotonicity of $f(x)$ in~\eqref{eq:StoR}.}
\end{figure}
By starting from a stable solution for a given $\lambda$ and changing the parameters $h$ and $q$ along some line until the stable solution ceases to exist (which can be determined to a high precision using bisections) a phase diagram, mapping the stability regions of two types of hopfions, was computed and is presented in Fig.~\ref{fig:phaseDiagram}. It is superimposed onto a well known phase diagram of the ground states of helimagnet~\cite{BH1994}, shown by the dashed lines. In all these ground states the magnetostatic energy is exactly zero, except for the case of skyrmions, where it is negligible. That's why the dashed lines do not depend on $\lambda$. Since, unlike skyrmions, the hopfions contain substantial amount of magnetic charges, the dependence of their stability region on $\lambda$ can't be ignored. Besides the case of negligible magnetostatics ($\lambda\rightarrow\infty$) a couple of particular illustrative values of $\lambda$ were used for either of the hopfion types, to show that they both can be stable when the magnetostatic interaction is present. In general, magnetostatic interaction makes hopfions less stable and increases their energy.

The inset in Fig.~\ref{fig:phaseDiagram} shows the real space profile of the equilibrium type 1 and type 2 hopfions for a specific set of $\uh$, $\uq$ and $\lambda$, where both of them can exist. Note also that, because the equilibrium $|\nu|$ is different, the type 2 hopfion's overall size $R$ is about $2.5$ times larger than that of type 1. The different sign of $\nu$ reflects the fact that the negative sign of $\uD$ is needed to turn the $\chi=3\pi/2$ DM energy maximum into the energy minimum. To have the insets show the hopfions in the same material with a particular fixed value of $\uD$, chirality of the trial function for the type 2 hopfion in the inset was reversed by plotting $w=-e^{\imath\chi}u^n/\overline{v^m}$ instead of~\eqref{eq:Whitehead}. This reversal of chirality has no effect on the energy of the hopfion, provided one also  changes the sign of $\uD$ and, consequently, $\nu$.

The narrow (although non-negligible and comparable e.g. to that of cones) stability region of the hopfions of the second type is the result of the requirement of monotonicity of $f(x)$. Should this requirement be lifted, the stability region expands considerably, as shown by the dotted line in Fig.~\ref{fig:phaseDiagram}. The case of non-monotonous $f(x)$ corresponds to a markedly different topology of the pre-images of the magnetization vectors. With monotonous $f$ each of the pre-images is a closed loop with different ones being interlinked exactly $\uHopf$ times. Non-monotonous $f$ makes some of the pre-images consist of two (or more) disconnected closed loops. Such states turn out to be more stable than a pure type 2 hopfions, but may belong to a different topological class.

Concluding, it is demonstrated here analytically that in a bulk helimagnet, described by the standard micromagnetic hamiltonian, two types of localized and topologically equivalent $\uHopf=1$ hopfion stable states exist. Numerical computation of the phase diagram shows that they may also coexist in the same material under the same conditions. These two types of hopfions are different by a constant rotation of all the magnetization vectors. They have different size, spatial profile, energies and stability, but are otherwise topologically equivalent, being similar in this respect to the Bloch and N{\'e}el domain walls, known in classical magnetism.

\begin{acknowledgments}
The support of the Russian Science Foundation under the project RSF~21-11-00325 is gratefully acknowledged.
\end{acknowledgments}
\bibliography{klm_base}

\end{document}